\documentclass{article}

\usepackage{scml2025}

\usepackage{multicol}
\usepackage[utf8]{inputenc} 
\usepackage[T1]{fontenc}    
\usepackage{hyperref}       
\usepackage{url}            
\usepackage{booktabs}       
\usepackage{amsfonts}       
\usepackage{nicefrac}       
\usepackage{amsmath}
\usepackage{microtype}      
\usepackage{cleveref}       
\usepackage{lipsum}         
\usepackage{graphicx}
\usepackage{doi}
\usepackage{wrapfig}
\usepackage{float}
\usepackage{subcaption}     
\usepackage{algorithm}
\usepackage{algpseudocode}
\usepackage[numbers]{natbib}

\title{Regression-based Physics-informed Neural Network (\emph{Reg-PINN}) for Magnetopause Tracking}
\renewcommand{\shorttitle}{Reg-PINN for Magnetopause Tracking}

\date{}

\newif\ifuniqueAffiliation
\uniqueAffiliationtrue

\ifuniqueAffiliation 
\author{%
    Po-Han K.~Hou\thanks{Department of Space Science and Engineering, National Central University, Chungli, Taiwan} \\
    Department of Earth Science and Engineering \\ 
    Imperial College London \\
    London, SW11 2DL \\
    \texttt{pohan.hou24@imperial.ac.uk}\\
    \And
    Sung-Chi C.~Hsieh \\
    Department of Electrical Engineering \\
    University of Leicester \\
    Leicester, LE1 7RH \\
    \texttt{sch45@student.le.ac.uk} \\
}
\else
\usepackage{authblk} 

\author[1,2]{Po-Han K.~Hou}
\author[3]{Sung-Chi C.~Hsieh}
\affil[1]{Department of Earth Science and Engineering, Imperial College London, London, SW11 2DL}
\affil[2]{Department of Space Science and Engineering, National Central University, Chungli, Taiwan}
\affil[3]{Department of Electrical Engineering, University of Leicester, Leicester, LE1 7RH}

\fi

\hypersetup{
    pdftitle={Regression-based Physics-informed Neural Networks for Magnetopause Tracking},
    pdfsubject={physics-informed neural networks, magnetopause tracking},
    pdfauthor={Po-Han K.~Hou, Sung-Chi C.~Hsieh},
    pdfkeywords={physics-informed neural networks, empirical models, magnetopause tracking},
}

\begin{document}
\maketitle

\begin{abstract}
    Previous research in the scientific field has utilized statistical empirical models and machine learning to address fitting challenges. While empirical models provide numerical generalizations, they often sacrifice precision. Traditional machine learning methods can provide high accuracy but may lack necessary generalization. We present a regression-based physics-informed neural network (Reg-PINN) that integrates physics-inspired empirical models into the neural network's loss function, thus merging the advantages of generalization and precision. The study validates the proposed method using the magnetopause boundary location as the target and explores the feasibility of methods including Shue et al. [1998], a data overfitting model, a fully connected network, Reg-PINN with Shue's model, and Reg-PINN with the overfitting model. Compared to Shue's model, this technique achieves an approximately 30\% reduction in RMSE, presenting a proof-of-concept improved solution for the scientific community.
\end{abstract}


\begin{multicols}{2}

\section{Introduction}
Over the past 50 years, various theories have been developed to describe the pressure equilibrium boundary between Earth's magnetosphere and the solar wind, known as the magnetopause. These include an investigation of the pressure balance between the magnetosphere and solar wind in 1969 \citep{schield1969} and an empirical model for the magnetopause proposed in 1997 and 1998 \citep{shue1997, shue1998}, which has been widely acknowledged as the leading approach for the past 25 years \citep{sws_satellite_observations}. Additionally, a method using support vector regression (SVR) was introduced in \citep{wang2013} to estimate the magnetopause position.

Traditional empirical models excel in generalization but sacrifice precision, while machine learning models achieve high precision during training but may lack the generalization capability of empirical models. Thus, this study aims to combine the strengths of both approaches by proposing a novel fitting algorithm called Regression-based Physics-Informed Neural Networks (Reg-PINN).

Reg-PINN are inspired by Physics-Informed Neural Networks (PINNs), which have been used for high-resolution time evolution or spatial state simulations of physical phenomena governed by differential equations. This study introduces a new approach that utilizes a regression-based equation form for training PINNs, combining physics-inspired fitting with neural networks to enhance both precision and generalization. The proposed method was tested against the widely used empirical model in \citep{shue1998} as a baseline, achieving approximately a 30\% reduction in root mean square error (RMSE).

\section{Data Collection and Preprocessing}

\begin{figure}[H]
    \centering
    \begin{subfigure}[b]{0.495\columnwidth}
        \centering
        \includegraphics[width=\textwidth]{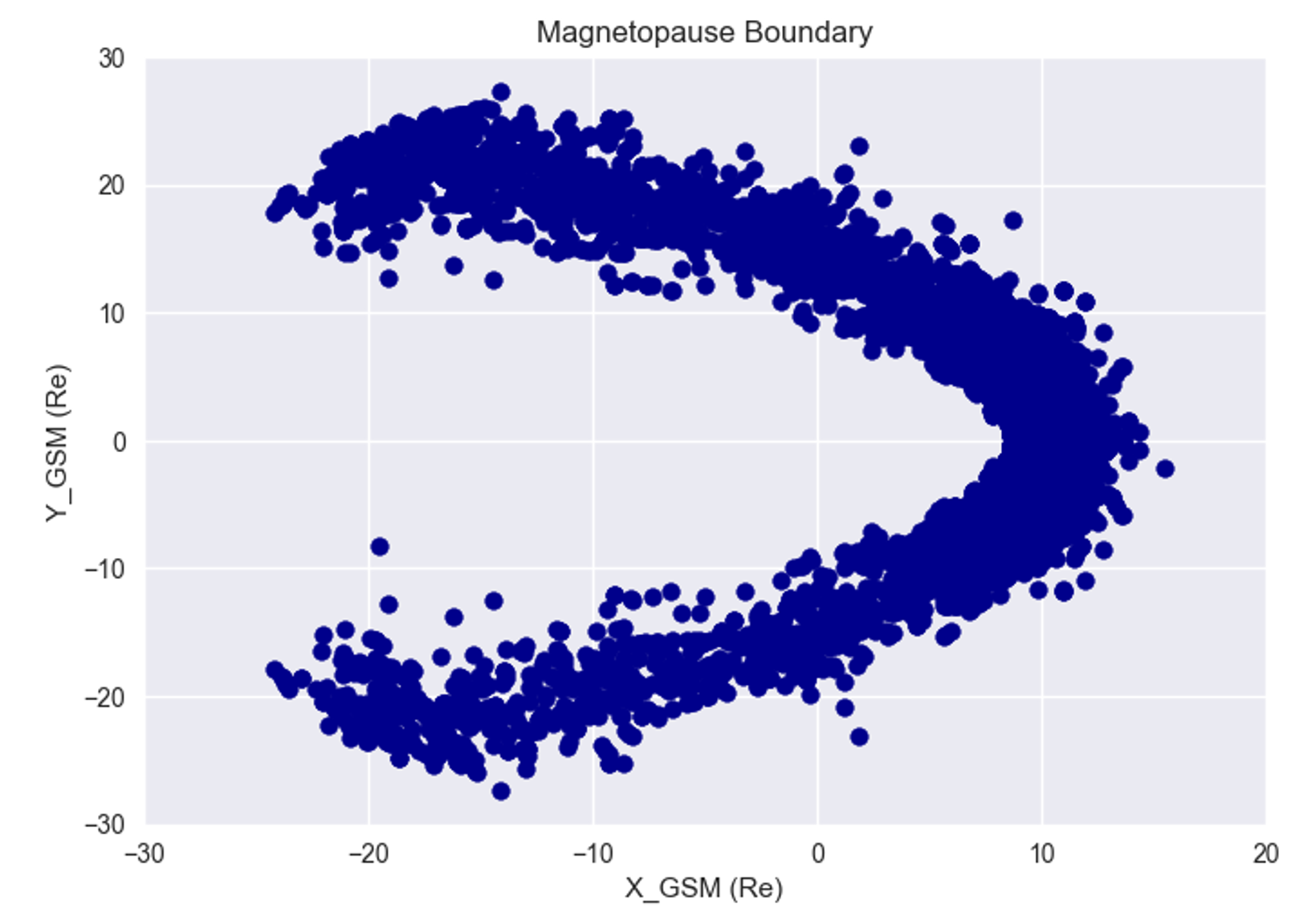}
        \caption{}
        \label{fig:mp_data}
    \end{subfigure}
    \hfill
    \begin{subfigure}[b]{0.495\columnwidth}
        \centering
        \includegraphics[width=\textwidth]{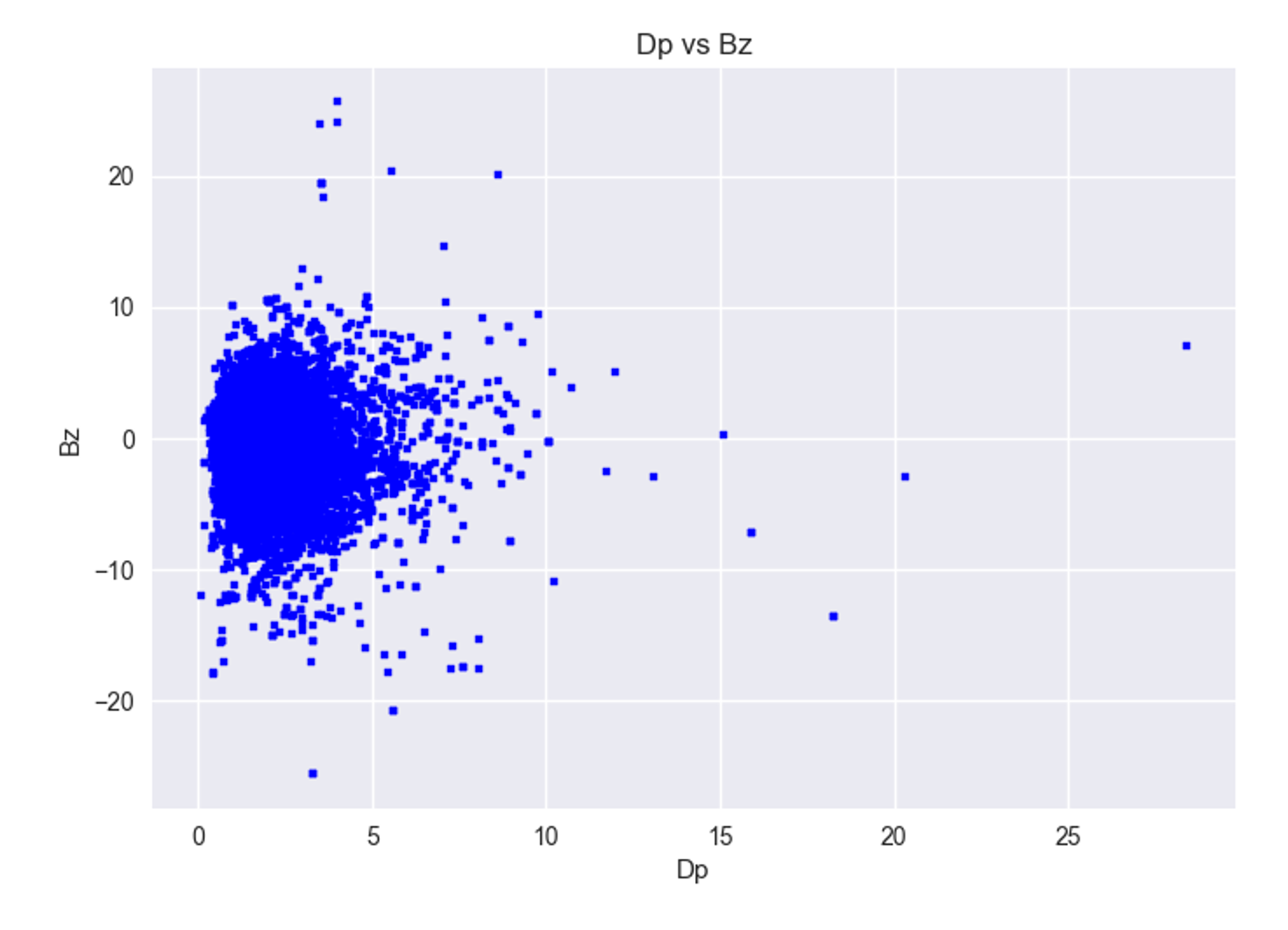}
        \caption{}
        \label{fig:mp_scatter}
    \end{subfigure}
    \caption{Magnetopause crossing data visualization, including (a) distribution of crossing data for origin is the Earth position and (b) $D_p$ v.s. $B_z$ distribution.}
    \label{fig:data_visualization}
\end{figure}

A total of 34,998 magnetopause in-situ crossing data points were derived from NASA OMNIWeb Data Explorer \citep{nasa_omniweb}, including THEMIS (28,634) \citep{themis_overview_data}, Geo-Tail (5,764), ISEE, and IMP 8. In Figure \ref{fig:data_visualization}, the inverted C-shaped data pattern represents the in-situ crossing data points under GSM coordinates and illustrates the distribution of the north-south magnetic field and the solar wind dynamic pressure ($D_p$). For each data point, the corresponding timestamps were identified in the 5-minute averages and integrated with solar wind ion flux, density, composition, and flow pressure recorded at the same time intervals for its consistency. To examine the model's generalizability within a specific region, we integrated the processed data for each bin.

Each bin encompasses north-south interplanetary magnetic field (IMF $B_z$) values spanning 3 nT, with a shift of 1 nT for each subsequent bin. For $D_p$, each bin includes two values with a shift of 0.5 nPa for each subsequent bin. The dataset range used in this study is consistent with that in \citep{shue1998}, specifically $-18$ nT < $B_z$ < $15$ nT and $0.5$ nPa < $D_p$ < $8.5$ nPa. The data preprocessing stage ensured the dataset was suitable for training and comparison with empirical models.

\section{Methodologies}
There have been numerous studies discussing the construction of empirical models for magnetopause location, including notable works in 1993 \citep{roelof1993}, 1997, 1998 \citep{shue1997, shue1998}, and 2002 \citep{chao2002}. In this study, we propose the use of Reg-PINN, which aims to combine the strengths of both empirical models and machine learning methods by achieving high accuracy and generalization capabilities.

\subsection{Physics-Inspired Empirical Model}
The position of the magnetopause is predominantly determined by $B_z$ and $D_p$ \citep{roelof1993, lu1994, fairfield1979}. Consequently, Shue employs data from satellites such as ISEE, AMPTE/IRM, and IMP 8, utilizing statistical analysis of these factors to ascertain the radial distance $r$ from the Earth-Sun line at an angle $\theta$, measured in Earth radii (Re). The fitting function of the magnetopause boundary, as presented in this study, adopts the formulation proposed by Shue \citep{shue1997} under polar coordinates, where $r_0$ is the standoff distance between Earth and the subsolar point of the magnetopause, and $\alpha$ represents the tail flaring, as expressed in Equation \ref{eq:shue_model}, and utilizes the parameters from \citep{shue1998} as the baseline, as shown in Equation \ref{eq:shue_model_parameters}. 

Figure \ref{fig:r0_bz} illustrates a significant variation when $B_z$ 
transitions from -10 nT to 0 nT. This phenomenon is explained by the 
southward IMF reconnecting on the dayside, subsequently compressing and pushing towards the magnetopause in the direction of Earth \citep{cassak2016reconnection, sonnerup1984magnetic}. That is the reason why Shue imports a hyperbolic tangent function to fit the curve to follow the theorem.

\footnotesize
\begin{align}
    \label{eq:shue_model}
    r &= r_0 \left(\frac{2}{1 + \cos(\theta)}\right)^{\alpha}
\end{align}

\begin{equation}
    \label{eq:shue_model_parameters}
    \left\{
    \begin{array}{ll}
        r_0(B_z, D_p) &= 10.22 + 1.29 \cdot \tanh\left(0.184 \cdot (B_z + 8.14)\right) ({D_p})^{\frac{1}{6.6}}, \\
        \alpha(B_z, D_p) &= (0.58 - 0.007 B_z) [1 + 0.024 \ln(D_p)]
    \end{array}
    \right.
\end{equation}

\subsection{Data-Overfitting Model}
\begin{figure}[H]
    \centering
    \begin{subfigure}[b]{0.495\columnwidth}
        \centering
        \includegraphics[width=\textwidth]{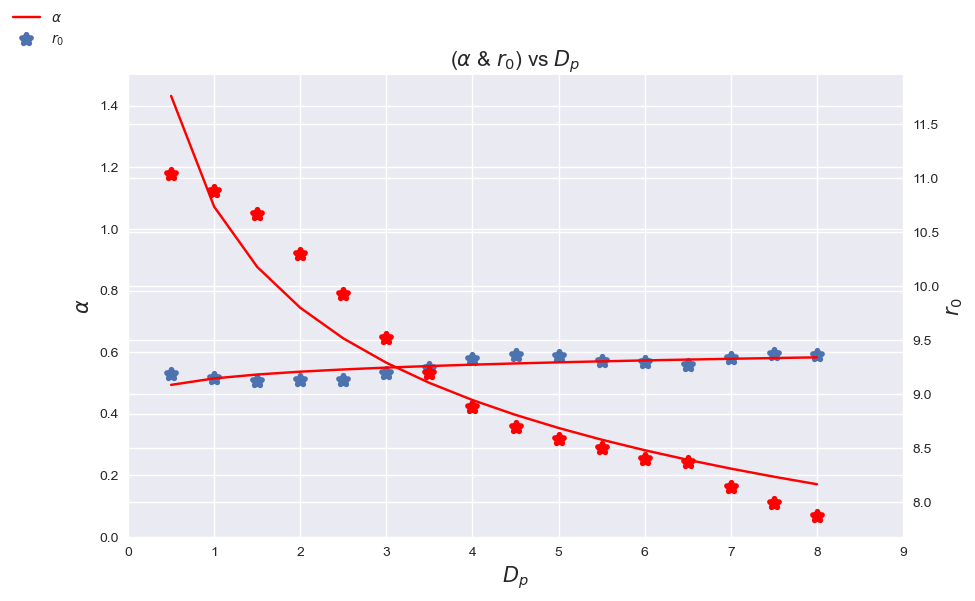}
        \caption{}
        \label{fig:r0_dp}
    \end{subfigure}
    \hfill
    \begin{subfigure}[b]{0.495\columnwidth}
        \centering
        \includegraphics[width=\textwidth]{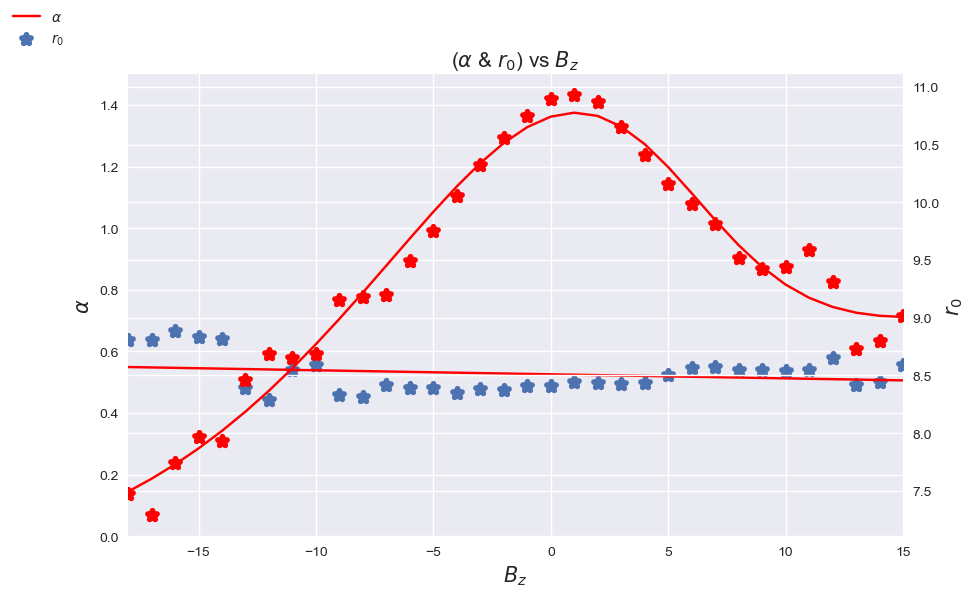}
        \caption{}
        \label{fig:r0_bz}
    \end{subfigure}
    \caption{Parameters for fitting the empirical model. (a) ($\alpha$ \& $r_0$) v.s. $D_p$, and (b) ($\alpha$ \& $r_0$) v.s. $B_z$.}
    \label{fig:empirical_model_parameters}
\end{figure}

 The most significant difference in parameter fitting compared to the original formulation in \citep{shue1998} is the decay of $r_0$ relative to $B_z$ when $B_z$ > 0 nT, as shown in Figure \ref{fig:r0_bz}. This phenomenon was not observed in Shue's dataset, which led \citep{shue1998} to utilize a single hyperbolic tangent function for curve fitting. However, using the same THEMIS satellite dataset, studies in \citep{nemecek2010} and \citep{li2023} have observed this phenomenon. Consequently, we have reevaluated and modified the original parameter formulation to employ two hyperbolic tangent functions, despite the lack of foundational theoretical support. We performed individual fitting of each variable using the least squares method as in \citep{shue1997} with the result shown in Equation \ref{eq:modified_shue}. It is worth noting that the power law relationship between $D_p$ and $r_0$, as mentioned by Schield in 1996 \citep{schield1969}, refers to the case of an ideal dipole magnetic field in a vacuum, where the relationship between the magnetopause boundary position $r_0$ and the solar wind dynamic pressure $D_p$ should follow a negative exponential power law with an exponent of 1/6. This value serves as a reference for comparing the actual relationship between $r_0$ and $D_p$ under realistic conditions. Shue, on the other hand, calculated an exponent of 1/6.6 in 1998 \citep{shue1998}, while in this study, the calculated exponent is 1/6.22.

\begin{equation}
   \label{eq:modified_shue}
   \left\{
   \begin{array}{ll}
       r_0 &= \left( 9.332 + 1.308 \cdot \tanh \left(0.213(B_z + 11.191)\right) \right. \\
       &\qquad \left. - 0.568 \cdot \tanh \left(0.479(B_z - 7.188)\right) \right)\left(D_p\right)^{\left(\frac{-1}{6.22}\right)}, \\
       \alpha &= \left( 0.493 - 3.5 \times 10^{-4} \cdot B_z \right)\left(D_p\right)^{\left(\frac{1}{11.92}\right)}
   \end{array}
   \right.
\end{equation}

\subsection{Regression-based Physics Informed Neural Networks}
Neural networks have been used for fitting problems for many years, with the Universal Approximation Theorem \citep{chen1995} being the main theoretical support for using machine learning in fitting tasks. The primary objective of implementing neural networks is to project a high-dimensional manifold of features onto the desired solution. The term "universal" signifies that regardless of the mathematical expression $y = f(\theta)$ that needs to be solved, or the pattern of $f$, neural networks can approximate a solution model that conforms to the desired spatial features. In other words, it can approximate any continuous function to arbitrary accuracy on a compact input domain.

Let $f(\theta)$ be a continuous function defined on a compact subset $K \subseteq \mathbb{R}^n$. Given any $\epsilon > 0$, there exists a neural network function $NN(\theta)$ such that

\begin{align}
    \sup_{\theta \in K} |f(\theta) - NN(\theta)| < \epsilon
\end{align}

The fitting for the $NN(\theta)$ consists of multiple non-linear and linear combinations with respect to a controllable cost function to optimize the model parameters or grant a discipline to follow the desired pattern. We used a regression pattern based on data distribution that was inspired by physics and added an external term for the loss function called $L_{reg}$. This makes the total loss function $L_{total}$ used in Reg-PINN as follows:

\begin{equation}
    L_{total} = L_{data} + \lambda L_{reg}
\end{equation}



where $L_{data}$ represents the mean square error (MSE) between the predicted values and the known data points, $L_{reg}$ stands for the MSE between the regression model output and the network output (serving as a regularization term with a desired pattern), and $\lambda$ is a hyperparameter that acts as a weighting parameter for the desired importance of the regression model. In the experiment, $\lambda = 1$ was used.

\begin{algorithm}[H]
\small
\caption{Reg-PINN}
\label{alg:reg_pinns}
\begin{algorithmic}[1]
    \State \textbf{Input:} Dataset $D = \{ [x_1, x_2, x_3, \dots, x_n] \rightarrow [y_1, y_2, y_3, \dots, y_m] \}$
    \State \textbf{Output:} Regularized model obtained via the empirical formula (regression form)
    \State Set $\mathbf{Y}_{\text{reg}} = f_{\text{reg}}(\mathbf{X})$.
    \State Set maximum iteration $(K)$, threshold for model $(\epsilon_{\text{threshold}})$, learning rate $(\eta)$, weighted parameters $(\lambda)$.
    \State Set $NN(\mathbf{X})$.
    \State \textbf{Initialize:} iteration = 0, $\epsilon = \infty$.
    \While{$\epsilon > \epsilon_{\text{threshold}}$ and iteration $< K$}
        \State $\mathbf{Y}_{\text{NN}} = NN(\mathbf{X})$
        \State $L_{\text{NN}}: L_{\text{NN}} = \frac{1}{m}\sum_{i=1}^{m} (y_{\text{NN}_i} - y_i)^2$
        \State $L_{\text{reg}}: L_{\text{reg}} = \frac{1}{m}\sum_{i=1}^{m} (y_{\text{reg}_i} - y_i)^2$
        \State $L_{\text{total}} = L_{\text{NN}} + \lambda L_{\text{reg}}$
        \State $W_i \leftarrow W_i - \eta \cdot \frac{\partial L_{\text{total}}}{\partial W_i}$
        \State $\epsilon = L_{\text{total}}$
        \State iteration = iteration + 1
    \EndWhile
\end{algorithmic}
\end{algorithm}

\section{Results and Discussions}
\begin{figure}[H]
    \centering
    \begin{subfigure}[b]{0.495\columnwidth}
        \centering
        \includegraphics[width=\textwidth]{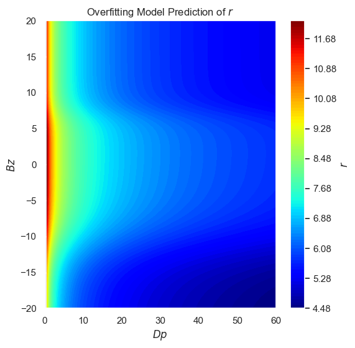}
        \caption{}
        \label{fig:overfitting}
    \end{subfigure}
    \hfill
    \begin{subfigure}[b]{0.495\columnwidth}
        \centering
        \includegraphics[width=\textwidth]{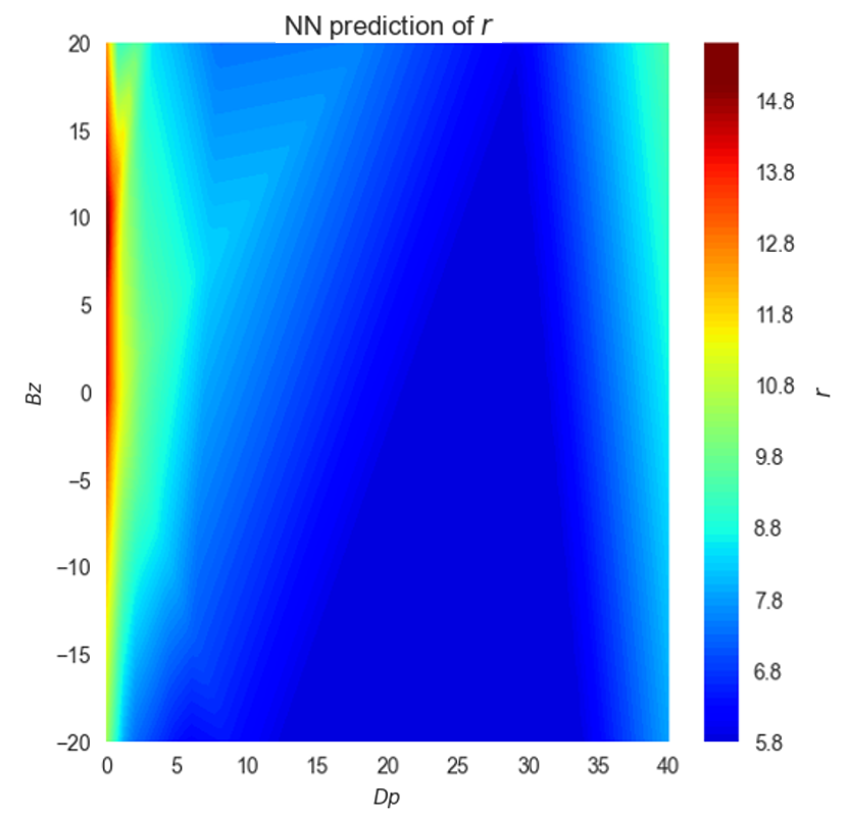}
        \caption{}
        \label{fig:nn}
    \end{subfigure}
    \hfill
    \begin{subfigure}[b]{0.495\columnwidth}
        \centering
        \includegraphics[width=\textwidth]{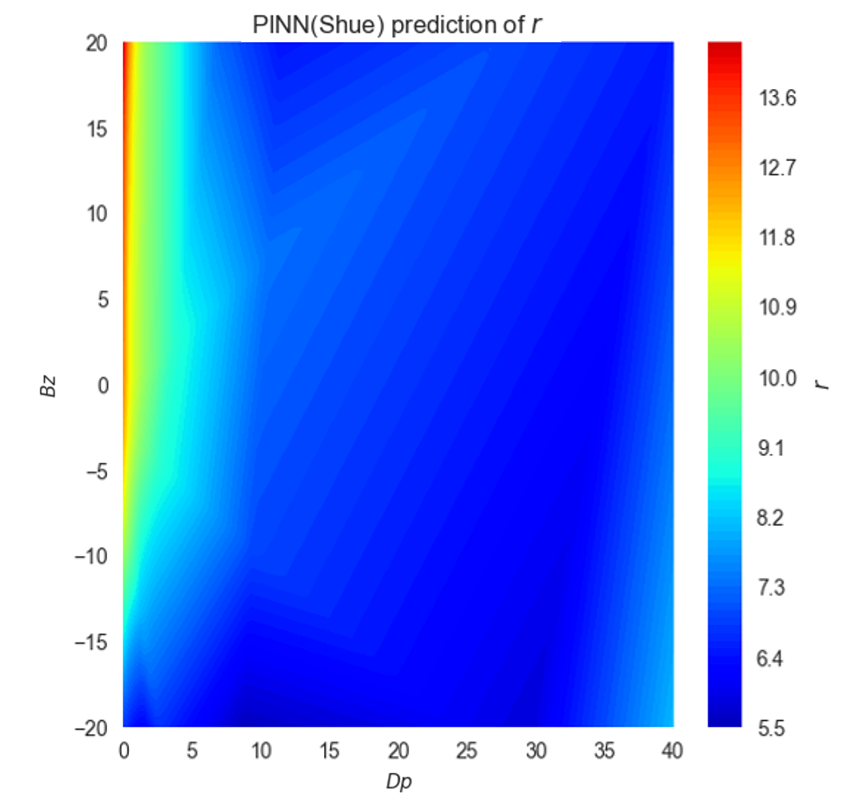}
        \caption{}
        \label{fig:pinn_shue}
    \end{subfigure}
    \hfill
    \begin{subfigure}[b]{0.495\columnwidth}
        \centering
        \includegraphics[width=\textwidth]{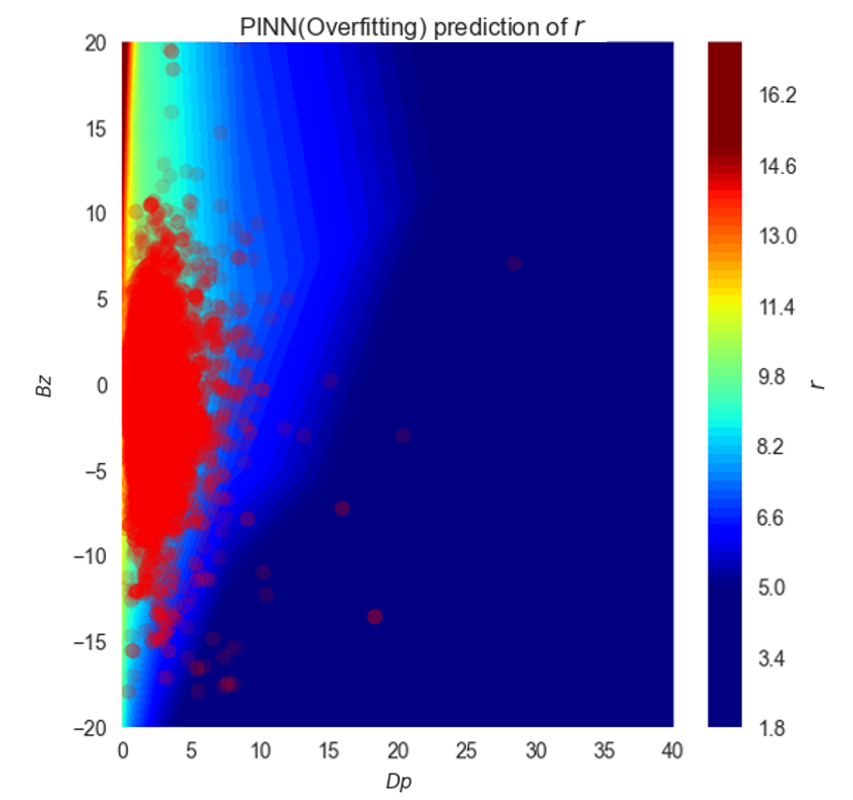}
        \caption{}
        \label{fig:pinn_of}
    \end{subfigure}
    \caption{Contour plot of different models, including (a) overfitting, (b) Vanilla NN, (c) Reg-PINN (Shue), and (d) Reg-PINN (Overfitting) along with ($D_p$ v.s. $B_z$) distribution for comprehensively visualization.} 

    \label{fig:result_vizualization}
\end{figure}

The experimental results are shown in Figure \ref{fig:result_vizualization}. The contour plot of the overfitting model in Figure \ref{fig:overfitting} and the pure machine learning model in Figure \ref{fig:nn} exhibit a bulged pattern under $D_p < 10$ nPa in the contour plot. In contrast, the Reg-PINN (Shue), which inherits Shue's physics-inspired constraints, stabilises when $B_z > 0$ nT (see \nameref{appendix}). 

The performance of the proposed Reg-PINN was evaluated using root mean square error (RMSE) as the primary metric. Table \ref{tab:rmse_num} shows the RMSE for the fitting based on empirical analysis, using Shue's model as the baseline, an overfitted model, and an overfitted model with parameters sampled using Markov Chain Monte Carlo (MCMC). These models were tested with both the full set of 34,998 data points and the dataset used in \citep{shue1998}. The results show that the statistical model provides only a marginal improvement, even with the modified overfitting approach.

\begin{table}[H] 
\centering
\caption{RMSE comparison of different empirical models}
\label{tab:rmse_num}
\begin{tabular}{lcc}
\hline
Model & RMSE (All) & RMSE (Shue) \\
\hline
Shue: Baseline & 1.316 $R_e$ & 1.347 $R_e$  \\
Overfitting (O.F.) & \textbf{1.192} $R_e$ & 1.297 $R_e$  \\
MCMC &  1.195 $R_e$ & \textbf{1.283} $R_e$  \\
\hline
\end{tabular}
\end{table}

\begin{figure}[H]
    \centering
    \begin{subfigure}[b]{0.495\columnwidth}
        \centering
        \includegraphics[width=\textwidth]{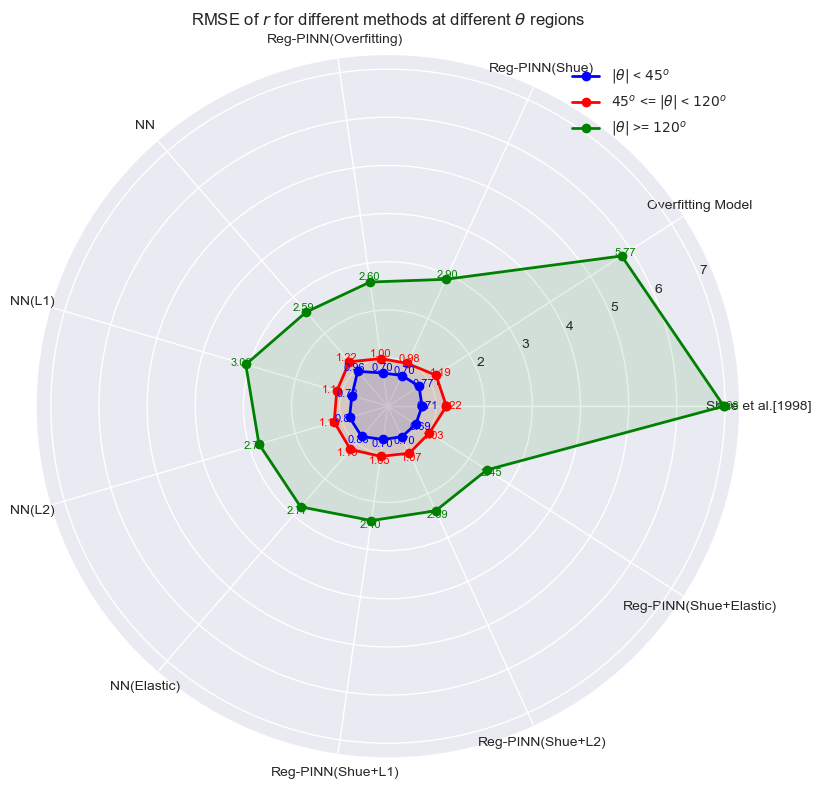}
        \caption{}
        \label{fig:rmse_theta}
    \end{subfigure}
    \hfill
    \begin{subfigure}[b]{0.495\columnwidth}
        \centering
        \includegraphics[width=\textwidth]{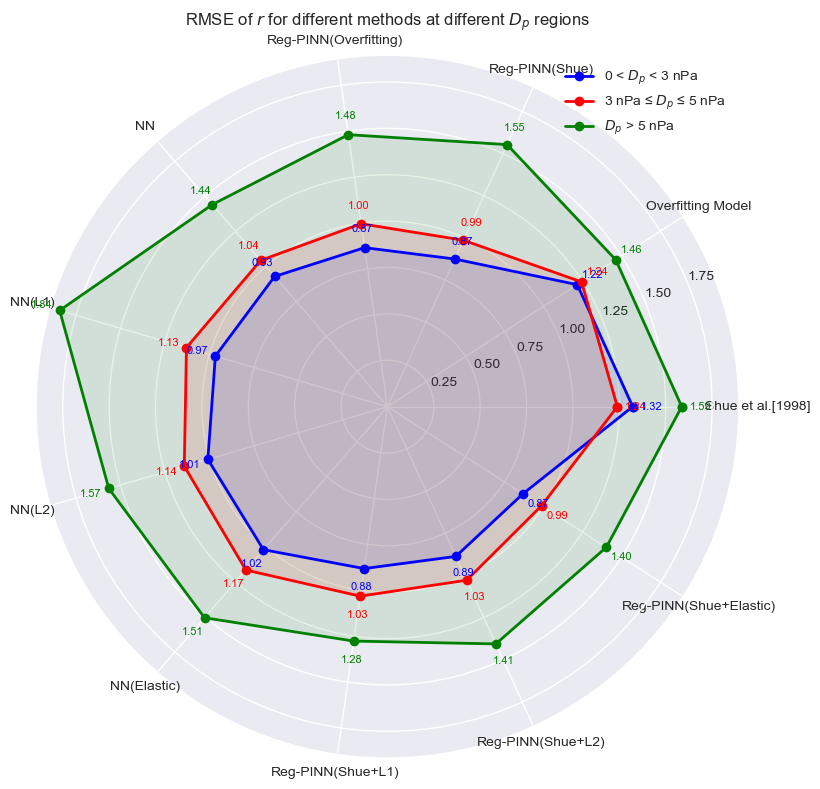}
        \caption{}
        \label{fig:rmse_dp}
    \end{subfigure}
    
    \vspace{0.5cm}
    
    \begin{subfigure}[b]{0.495\columnwidth}
        \centering
        \includegraphics[width=\textwidth]{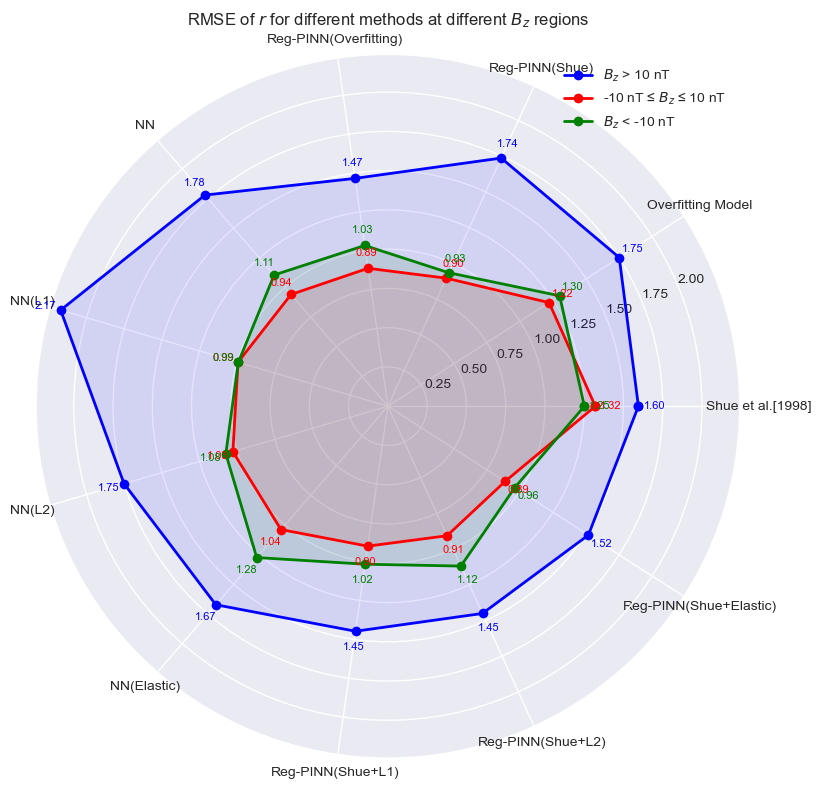}
        \caption{}
        \label{fig:rmse_bz}
    \end{subfigure}
    \hfill
    \begin{subfigure}[b]{0.495\columnwidth}
        \centering
        \includegraphics[width=\textwidth]{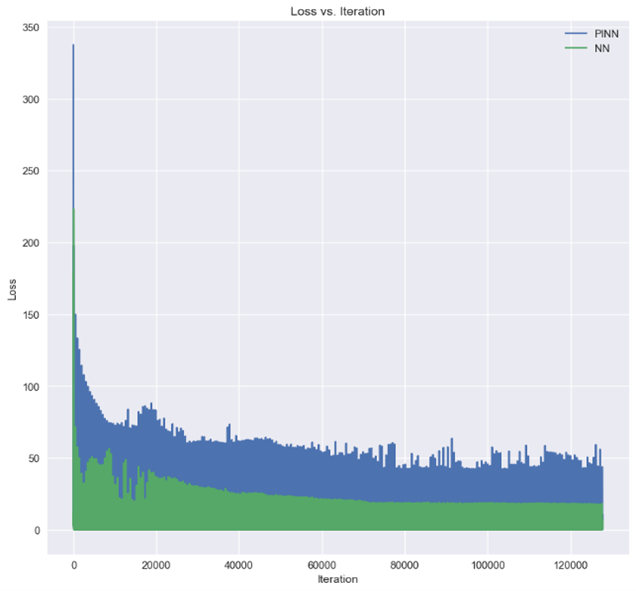}
        \caption{}
        \label{fig:loss}
    \end{subfigure}
    \caption{RMSE for evaluating different physical parameters, including (a) $\theta$, (b) $D_p$, and (c) $B_z$, along with (d) Loss vs. Iteration for NN and Reg-PINN on this case.}
    \label{fig:rmse_parameters}
\end{figure}

Table \ref{tab:rmse_nn} compares the ML-based method to the baseline, divided into two groups: the first column shows results with 80\% of data used for training and 20\% masked for testing, while the second column shows results with 20\% of data used for training and 80\% masked for testing. Each experiment was structured with [3 x 27 x 81 x 27 x 9 x 1] layers, an RMSProp optimizer, and 500 epochs. Figure \ref{fig:loss} illustrates that Reg-PINN's loss remains consistently higher than NN's loss. The vanilla NN achieved the best performance with 80\% of the data for training, but it lost precision when trained on a smaller dataset. In contrast, the Reg-PINN demonstrated consistent improvement in capturing the magnetopause position across varying data amounts, indicating its generalizability and precision.

\begin{table}[H]
\centering
\caption{RMSE comparison of different ML models}
\label{tab:rmse_nn}
\begin{tabular}{lcc}
\hline
Model & RMSE (20\%) & RMSE (80\%) \\
\hline
Baseline & 1.243 $R_e$ & 1.266 $R_e$  \\
Vanilla NN & \textbf{0.866} $R_e$ & 1.073 $R_e$  \\
Reg-PINN (Shue) & 0.889 $R_e$ & 0.932 $R_e$  \\
Reg-PINN (O.F.) & 0.873 $R_e$ & \textbf{0.918} $R_e$ \\
\hline
\end{tabular}
\end{table}

Figure \ref{fig:rmse_parameters} presents the performance of each test model: Shue's 1998 model, the overfitting model, Reg-PINN (Shue, overfitting, Shue-L1, Shue-L2, and Shue-Elastic), vanilla neural networks (L1, L2, and Elastic), and under varying $\theta$, $D_p$, and $B_z$ parameters, with 50\% of the data used for training and 50\% reserved as unseen data for testing.

In most parameter regions, the ML-based method performs competitively with the conventional approach, as illustrated in Figure \ref{fig:rmse_dp} for $D_p$ $\leq$ 5 nPa and Figure \ref{fig:rmse_bz} for $B_z$ $\leq$ 10 nT. However, the proposed method emphasizes the benefits of integrating neural networks with empirical models, as this enables the inclusion of low-probability events such as |$B_z$| > 10 nT, $D_p$ > 5 nPa, and |$\theta$| > 120$^o$, with probabilities of 0.62\%, 1.55\%, and 1.56\%, respectively. The most significant improvement for the ML-based approach is observed in Figure \ref{fig:rmse_theta} for the ML-based approach handling |$\theta$| > 120$^o$.
In figure \ref{fig:rmse_dp}, the result for $D_p$ > 5 nPa shows that the NN method performs the worst among the models, indicating potential data insufficiency under extreme conditions. With the help of \citep{shue1998}, which has considered extreme solar wind conditions, the Reg-PINN (Shue) can follow the underlying constraint, resulting in better predictions under machine learning optimization. This also indicates that the model can learn from the pattern of a physics-inspired model, providing a robust approach for acquiring the magnetopause position.

\begin{figure}[H]
    \centering
        \centering
        \includegraphics[width=0.495\textwidth]{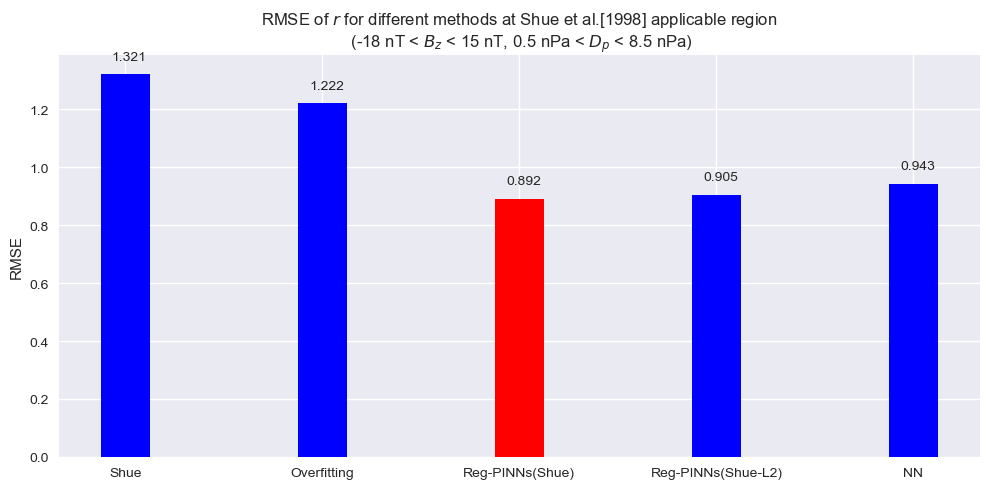}
        \caption{Selected model RMSE evaluated under the baseline applicable region}
        \label{fig:shue_rmse}
    \hfill
\end{figure}

Although the ML-based approach yields lower error, the research focuses on physics-inspired techniques supported by real physical governing principles for modeling. Under these circumstances, we present the overfitting model, Reg-PINN (Shue), Reg-PINN (Shue-L2), and a vanilla neural network to evaluate performance, as shown in Figure \ref{fig:shue_rmse}. The decision to use Reg-PINN with Shue's model is based on its alignment with the underlying theoretical framework, offering a more "physics-informed" perspective with an error reduction rate of 32.5\% compared to Shue's model. While numerous new machine learning algorithms can achieve high accuracy in fitting tasks, few are grounded in real physics. The experiment further utilize an overfitting model within Reg-PINN to provide evidence of Reg-PINN' robustness and precision over empirical or data-centric approaches when applied to new datasets, providing a simple proof of concept for enhancing applied scientific models with empirical foundations.

\begin{table}[H]
\centering
\caption{RMSE comparison of different $\lambda$ with Reg-PINN(Shue)}
\label{tab:rmse_lambda}
\begin{tabular}{lcc}
\hline
Model & RMSE (20\%) & RMSE (80\%) \\
\hline
Reg-PINN ($\lambda$ = 1) & \textbf{0.866} $R_e$ & 0.932 $R_e$  \\
Reg-PINN ($\lambda$ = 2) & 0.946 $R_e$ & 1.063 $R_e$ \\
Reg-PINN ($\lambda$ = 5) & 1.092 $R_e$ & 1.118 $R_e$ \\
Reg-PINN ($\lambda$ = 0.1) & 0.884 $R_e$ & \textbf{0.927} $R_e$  \\
Reg-PINN ($\lambda$ = 0.5) & 0.878 $R_e$ & 0.939 $R_e$  \\
\hline
\end{tabular}
\end{table}

\begin{figure}[H]
    \centering
    \begin{subfigure}[b]{0.495\columnwidth}
        \centering
        \includegraphics[width=\textwidth]{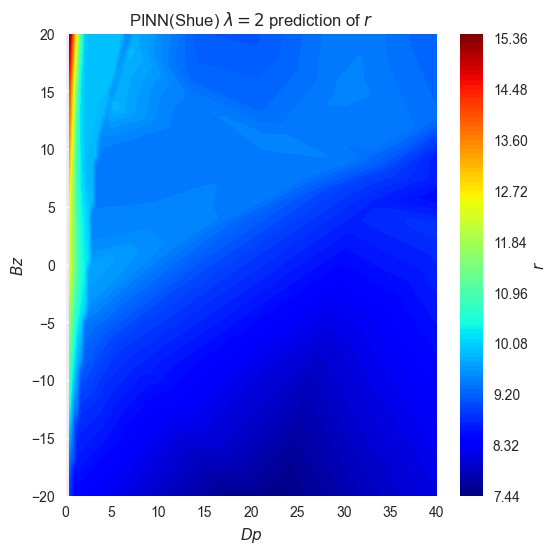}
        \caption{}
        \label{fig:lambda2}
    \end{subfigure}
    \hfill
    \begin{subfigure}[b]{0.495\columnwidth}
        \centering
        \includegraphics[width=\textwidth]{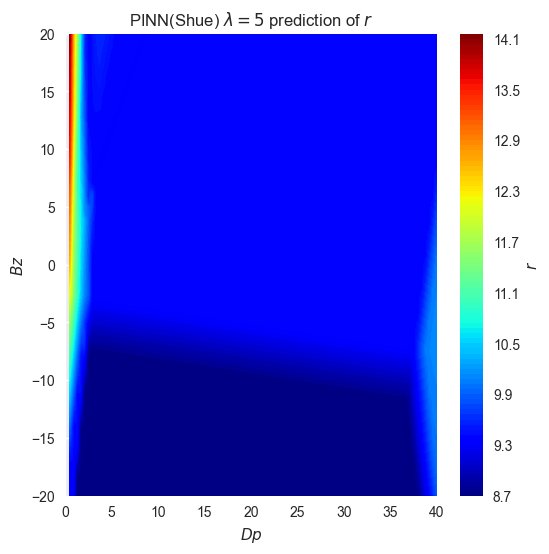}
        \caption{}
        \label{fig:lambda5}
    \end{subfigure}
    \hfill
    \begin{subfigure}[b]{0.495\columnwidth}
        \centering
        \includegraphics[width=\textwidth]{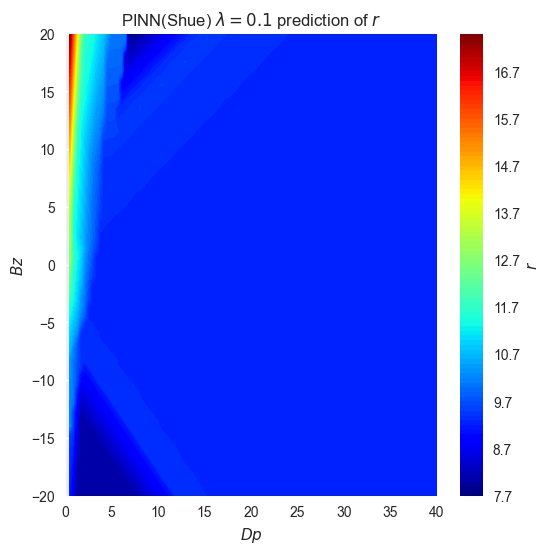}
        \caption{}
        \label{fig:lambda01}
    \end{subfigure}
    \hfill
    \begin{subfigure}[b]{0.495\columnwidth}
        \centering
        \includegraphics[width=\textwidth]{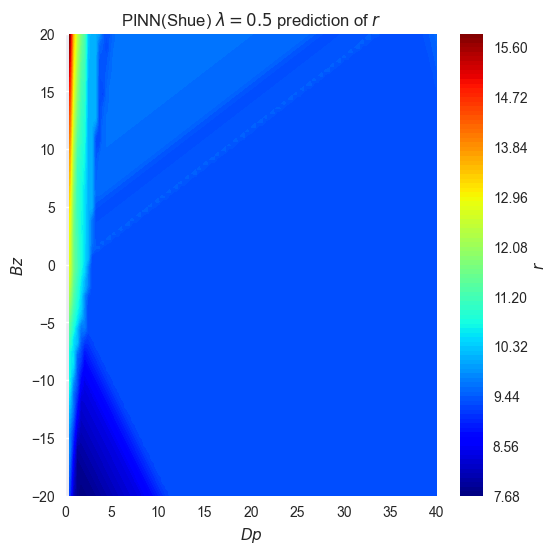}
        \caption{}
        \label{fig:lambda05}
    \end{subfigure}
    \caption{Contour plot for different $\lambda$ on Reg-PINN(Shue), including (a) $\lambda$ = 2, (b) $\lambda$ = 5, (c) $\lambda$ = 0.1, and (d) $\lambda$ = 0.5 along with ($D_p$ v.s. $B_z$) distribution for visualization.} 

    \label{fig:result_vizualization_lambda}
\end{figure}

\begin{figure}[H]
    \centering
        \centering
        \includegraphics[width=0.5\textwidth]{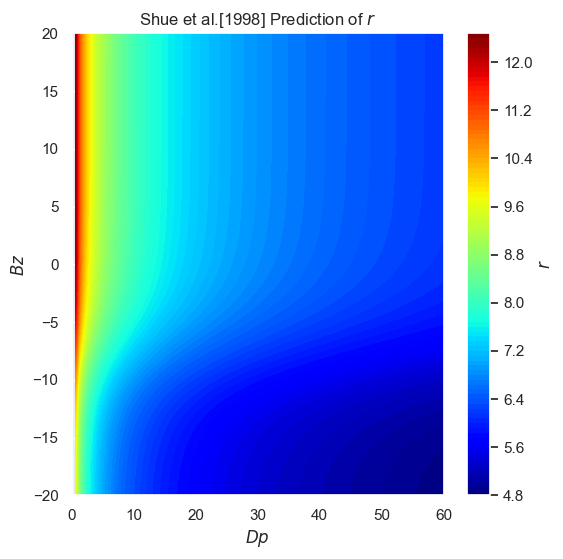}
        \caption{Shue's empirical model \citep{shue1998}}
        \label{fig:shue_model}
    \hfill
\end{figure}

Table \ref{tab:rmse_lambda} and Figure \ref{fig:result_vizualization_lambda} illustrate the different configurations of the weight ($\lambda$) for regression loss on Reg-PINN (Shue) alongside their corresponding contour plot, revealing distinct levels of influence from Shue's model as shown in Figure \ref{fig:shue_model}. It is worthy of mention that for the case of $\lambda$ = 2, 5, the results indicate a significant increase in RMSE compared to the cases of $\lambda$ = 0.1, 0.5. This pattern indicates that disproportionately prioritising the regression loss may result in an undue focus on data fitting, compromising the model's fundamental physical restrictions and perhaps causing numerical instability or overfitting to noise within the dataset. Consequently, the model exhibits poor generalisation, resulting in elevated RMSE values. This behaviour can be seen as an irreducible RMSE of the model, which seems to converge to about 0.86 $Re$ across different parameter configurations. This irreducible error presumably arises from constraints in the model's construction, the lack of a robust theoretical framework to adequately underpin the methodology, and the inherent uncertainties linked to measurement inaccuracies. In the scenario of evaluating 20\% masked data, the default value of $\lambda$ = 1 consistently yields the lowest RMSE. Conversely, when testing 80\% of the data with 20\% allocated for training, the minimum RMSE is attained with $\lambda$ = 0.1. However, in this instance, $\lambda \le 1$ yields comparable performance, whereas the performance declines as $\lambda$ increases.

\section{Conclusion}
The proposed algorithm, Reg-PINN, mitigates the precision constraints of empirical methods and the generalization difficulties in machine learning. Experimental evaluations focus on the feasibility of Reg-PINN with both a physics-inspired model and an overfitted model, demonstrating that Reg-PINN extends the capabilities of standard PINNs by incorporating algebraic equations into model training, extending the usual scope of ODEs/PDEs. Furthermore, the Reg-PINN forecasts the magnetopause position on unseen data with an error reduction of about 30\%. By enabling the integration of domain-specific empirical models with neural networks, this approach demonstrates potential for a wide range of scientific applications, such as remodeling the Fama-French three-factor model for stock returns \citep{FamaFrench1992} and understanding Hall-Petch strengthening in materials science \citep{Hansen2004, Cordero2016}. This flexibility allows for modeling and forecasting complex phenomena across disciplines, opening up new opportunities for scientific advancements and practical applications.

\section*{Acknowledgments}
There is no funding to be disclosed. The author would like to express our heartfelt appreciation to Dr. Chun-Yu Lin, associate researcher at the National Center for High-Performance Computing—National Applied Research Laboratories, for exchanging ideas and insights into the concept of PINNs and for engaging in fruitful discussions. In addition, we would like to extend our gratitude to Prof. Jih-Hong Shue, Professor at National Central University (NCU), for providing the data used in Shue et al. [1998], and to Mr. Yu-Wei Chen and Dr. Pai-Sheng Wang, researchers at the Space Environment Laboratory (NCU), for their contributions to reinforcing the understanding of the magnetopause.

\section*{Appendix: Contour plot of Shue's model }
\label{appendix}

\bibliographystyle{plain}
\bibliography{references}  

\begin{thebibliography}{10}

\bibitem{sws_satellite_observations}
{Australian Bureau of Meteorology, Space Weather Services}.
\newblock Satellite observations.
\newblock \url{https://www.sws.bom.gov.au/Satellite/3/2}.
\newblock Accessed: 2024-10-31.

\bibitem{cassak2016reconnection}
Paul~A. Cassak and Stephen~A. Fuselier.
\newblock Reconnection at earth's dayside magnetopause.
\newblock In Walter Gonzalez and Eugene Parker, editors, {\em Magnetic Reconnection}, volume 427 of {\em Astrophysics and Space Science Library}, pages 213--276. Springer, Cham, 2016.

\bibitem{chao2002}
J.K. Chao, J.-H. Shue, H.C. Fu, K.K. Khurana, C.T. Russell, and P.~Song.
\newblock Models for the size and shape of the earth’s magnetopause and bow shock.
\newblock In L.-H. Lyu, editor, {\em Space Weather Study Using Multipoint Techniques}, volume~12 of {\em COSPAR Colloquia Series}, pages 127--135. Pergamon, 2002.

\bibitem{chen1995}
Tianping Chen and Hong Chen.
\newblock Universal approximation to nonlinear operators by neural networks with arbitrary activation functions and its application to dynamical systems.
\newblock {\em IEEE Transactions on Neural Networks}, 6(4):911--917, 1995.

\bibitem{Cordero2016}
Zachary~C. Cordero, Braden~E. Knight, and Christopher~A. Schuh.
\newblock Six decades of the hall–petch effect – a survey of grain-size strengthening studies on pure metals.
\newblock {\em International Materials Reviews}, 61(8):495--512, 2016.

\bibitem{fairfield1979}
D.~H. Fairfield.
\newblock Structure of the magnetopause: Observations and implications for reconnection.
\newblock {\em Space Science Reviews}, 23(3):427--448, 1979.

\bibitem{FamaFrench1992}
Eugene~F. Fama and Kenneth~R. French.
\newblock The cross-section of expected stock returns.
\newblock {\em The Journal of Finance}, 47(2):427--465, 1992.

\bibitem{Hansen2004}
Niels Hansen.
\newblock Hall–petch relation and boundary strengthening.
\newblock {\em Scripta Materialia}, 51(8):801--806, 2004.

\bibitem{li2023}
Sheng Li, Yue-Ying Sun, and Chia-Hsien Chen.
\newblock An interpretable machine learning procedure which unravels hidden interplanetary drivers of the low latitude dayside magnetopause.
\newblock {\em Space Weather}, 21(3):e2022SW003391, 2023.

\bibitem{lu1994}
G.~Lu, L.~R. Lyons, and A.~D. Richmond.
\newblock Simulations of the polar ionosphere for varying interplanetary magnetic field conditions.
\newblock {\em Journal of Atmospheric and Terrestrial Physics}, 56(8):949--958, 1994.

\bibitem{nasa_omniweb}
{NASA Goddard Space Flight Center}.
\newblock {OMNIWeb Data Explorer}.
\newblock \url{https://omniweb.gsfc.nasa.gov/}.
\newblock Accessed: 2024-11-15.

\bibitem{nemecek2010}
Z.~Nemecek, J.~Safrankova, L.~Prech, E.~Grigorenko, J.~D. Richardson, and J.~Šafránková.
\newblock Magnetopause position under different conditions.
\newblock In {\em AGU Fall Meeting Abstracts}, pages SM13B--1812, 2010.

\bibitem{roelof1993}
E.~C. Roelof and D.~G. Sibeck.
\newblock Magnetopause shape as a bivariate function of interplanetary magnetic field bz and solar wind dynamic pressure.
\newblock {\em Journal of Geophysical Research: Space Physics}, 98(A12):21421--21450, 1993.

\bibitem{schield1969}
M.~A. Schield.
\newblock Pressure balance between solar wind and magnetosphere.
\newblock {\em Journal of Geophysical Research (1896-1977)}, 74(5):1275--1286, 1969.

\bibitem{shue1997}
J.-H. Shue, J.~K. Chao, H.~C. Fu, C.~T. Russell, P.~Song, K.~K. Khurana, and H.~J. Singer.
\newblock A new functional form to study the solar wind control of the magnetopause size and shape.
\newblock {\em Journal of Geophysical Research: Space Physics}, 102(A5):9497--9511, 1997.

\bibitem{shue1998}
J.-H. Shue, J.~K. Chao, H.~C. Fu, C.~T. Russell, P.~Song, K.~K. Khurana, and H.~J. Singer.
\newblock Magnetopause location under extreme solar wind conditions.
\newblock {\em Journal of Geophysical Research: Space Physics}, 103(A8):17691--17700, 1998.

\bibitem{sonnerup1984magnetic}
B.~U.~{\"O}. Sonnerup.
\newblock Magnetic field reconnection at the magnetopause: An overview.
\newblock In Edward W.~Jr. Hones, editor, {\em Magnetic Reconnection at the Magnetopause}, volume~30 of {\em Geophysical Monograph Series}, pages 92--99. American Geophysical Union, 1984.

\bibitem{themis_overview_data}
{University of California, Los Angeles, Institute of Geophysics and Planetary Physics}.
\newblock {THEMIS Overview Data}.
\newblock \url{https://themis.igpp.ucla.edu/overview_data.shtml}.
\newblock Accessed: 2024-10-31.

\bibitem{wang2013}
Y.~Wang, Y.~Zhou, Q.-G. Zong, H.~Zhang, P.~Song, L.~Xie, and S.~Fu.
\newblock A new three-dimensional magnetopause model with a support vector regression machine and a large database of multiple spacecraft observations.
\newblock {\em Journal of Geophysical Research: Space Physics}, 118(5):2173--2184, 2013.

\end{thebibliography}

\end{multicols}
\end{document}